\PassOptionsToPackage{unicode}{hyperref}
\PassOptionsToPackage{hyphens}{url}
\documentclass[
]{article}
\usepackage{lmodern}
\usepackage{amsmath}
\usepackage{ifxetex,ifluatex}
\ifnum 0\ifxetex 1\fi\ifluatex 1\fi=0 
  \usepackage[T1]{fontenc}
  \usepackage[utf8]{inputenc}
  \usepackage{textcomp} 
  \usepackage{amssymb}
\else 
  \usepackage{unicode-math}
  \defaultfontfeatures{Scale=MatchLowercase}
  \defaultfontfeatures[\rmfamily]{Ligatures=TeX,Scale=1}
\fi
\IfFileExists{upquote.sty}{\usepackage{upquote}}{}
\IfFileExists{microtype.sty}{
  \usepackage[]{microtype}
  \UseMicrotypeSet[protrusion]{basicmath} 
}{}
\usepackage{xcolor}
\IfFileExists{xurl.sty}{\usepackage{xurl}}{} 
\IfFileExists{bookmark.sty}{\usepackage{bookmark}}{\usepackage{hyperref}}
\hypersetup{
  pdftitle={Causally-Interpretable Random-Effects Meta-Analysis},
  pdfauthor={Justin M. Clark\^{}*; Kollin W. Rott\^{}*; James S. Hodges\^{}*; Jared D. Huling},
  hidelinks,
  pdfcreator={LaTeX via pandoc}}
\usepackage[margin=1in]{geometry}
\usepackage{graphicx}
\makeatletter
\def\maxwidth{\ifdim\Gin@nat@width>\linewidth\linewidth\else\Gin@nat@width\fi}
\def\maxheight{\ifdim\Gin@nat@height>\textheight\textheight\else\Gin@nat@height\fi}
\makeatother
\setkeys{Gin}{width=\maxwidth,height=\maxheight,keepaspectratio}
\makeatletter
\def\fps@figure{htbp}
\makeatother
\usepackage{setspace}
\usepackage[utf8]{inputenc}
\usepackage{silence}
\usepackage{bbm}
\usepackage{booktabs}
\usepackage{longtable}
\usepackage{array}
\usepackage{multirow}
\usepackage{wrapfig}
\usepackage{float}
\usepackage{colortbl}
\usepackage{pdflscape}
\usepackage{tabu}
\usepackage{threeparttable}
\usepackage{threeparttablex}
\usepackage[normalem]{ulem}
\usepackage{makecell}
\usepackage{xcolor}
\ifluatex
  \usepackage{selnolig}  
\fi
\newlength{\cslhangindent}
\newlength{\csllabelwidth}
\newenvironment{CSLReferences}[2] 
 {
  \setlength{\parindent}{0pt}
  \ifodd #1 \everypar{\setlength{\hangindent}{\cslhangindent}}\ignorespaces\fi
  \ifnum #2 > 0
  \setlength{\parskip}{#2\baselineskip}
  \fi
 }%
 {}
\usepackage{calc}

\title{Causally-Interpretable Random-Effects Meta-Analysis}
\author{Justin M. Clark\(^*\) \and Kollin W. Rott\(^*\) \and James S.
Hodges\(^*\) \and Jared D. Huling\footnote{University of Minnesota
  School of Public Health, Division of Biostatistics}}
\date{}

\begin{document}
\maketitle
\begin{abstract}
Recent work has made important contributions in the development of
causally-interpretable meta-analysis. These methods transport treatment
effects estimated in a collection of randomized trials to a target
population of interest. Ideally, estimates targeted toward a specific
population are more interpretable and relevant to policy-makers and
clinicians. However, between-study heterogeneity not arising from
differences in the distribution of treatment effect modifiers can raise
difficulties in synthesizing estimates across trials. The existence of
such heterogeneity, including variations in treatment modality, also
complicates the interpretation of transported estimates as a generic
effect in the target population. We propose a conceptual framework and
estimation procedures that attempt to account for such heterogeneity,
and develop inferential techniques that aim to capture the accompanying
excess variability in causal estimates. This framework also seeks to
clarify the kind of treatment effects that are amenable to the
techniques of generalizability and transportability.

\par

\end{abstract}

\newpage

\hypertarget{introduction}{%
\section{Introduction}\label{introduction}}

Given data from a collection of randomized controlled trials (RCTs), an
important question faced by clinicians and policy-makers alike is
whether such results apply to target populations of interest. Recent
work has made important advances in tackling this question by developing
causal inference methods designed for meta-analysis
(\protect\hyperlink{ref-dahabreh_biometrics}{Dahabreh et al. 2022}).
These techniques are designed to account for differences between the
trial and target populations, resulting in effect estimates that have a
causal interpretation for the target population.

The impact of such population differences is one example of a challenge
in research synthesis long recognized by practitioners of meta-analysis:
between-study heterogeneity of many kinds should be taken into account
when evaluating results from one trial to the next. Differences in the
characteristics of populations represents just one form of between-study
heterogeneity. In this paper, we develop causal quantities and
estimators that build on the existing causally-interpretable
meta-analysis framework to additionally take into account unexplained
between-study heterogeneity beyond that induced by covariate differences
between the collection of trials and the target population.

Our approach produces effect estimates that both are applicable to a
target population of scientific interest and remain interpretable even
when between-study heterogeneity prevents data pooling across trials. We
also see our work as a conceptual bridge between the developing field of
causally-interpretable meta-analysis and random-effects approaches that
are well-established in evidence synthesis. This structural resemblance
to traditional random-effects meta-analysis allow us to adopt some of
the analytical frameworks important to that approach while retaining
causal interpretability.

In principle, systematic reviews of randomized trials should have
important implications for health policy and clinical practice
(\protect\hyperlink{ref-jama_meta_analysis}{Berlin and Golub 2014}). For
example, the Strength of Recommendation Taxonomy (SORT) assigns
meta-analysis to the highest level of study quality
(\protect\hyperlink{ref-sort_taxonomy}{Ebell et al. 2004}). However,
recent work has suggested that meta-analyses--particularly meta-analyses
involving individual patient data (IPD)--have had a relatively small
impact on organizations developing and publishing clinical guidelines
(\protect\hyperlink{ref-ipd_meta_uptake}{Vale et al. 2015}), even among
guidelines for which relevant IPD meta-analyses were readily available.

Analogous work published in 2021 also found limited use of systematic
reviews to inform clinical guidelines: in that study, only 34\% of
analyzed guidelines conducted a systematic review of available evidence
to inform clinical practice
(\protect\hyperlink{ref-clinical_practice_bad}{Lunny et al. 2021}).
These authors give many possible explanations for their findings,
including ignorance on the part of guideline development groups, a
related over-reliance on expert opinion, and the amount of labor and
time required to complete a high-quality systematic review. However, it
is worth interrogating whether the low uptake of systematic reviews to
inform clinical guidelines might arise from the usefulness of the
meta-analyses themselves. That is, a review of clinical evidence may be
systematic, rigorous, and follow international standards while still
being of limited applicability to decision-making in a particular
population or setting.

Much of the relevance of a given trial to clinical practice stems from
how well the patient population targeted by the practitioner aligns with
trial participants. The complex, idiosyncratic process of recruitment
can result in studies that, while exhibiting a high level of internal
validity, are unlikely to apply to individuals distinct from trial
participants, thereby reducing external validity
(\protect\hyperlink{ref-generalizability}{Degtiar and Rose 2021}).
Concerns over external validity are especially relevant to health system
decision making, where beneficiaries may differ in key ways from
participants in the RCTs that are used to justify coverage decisions.
For example, a 2008 study suggested that participants in the RCTs
underlying a meta-analysis informing CMS policy differed substantially
from the Medicare population
(\protect\hyperlink{ref-cms_rct_variation}{Dhruva 2008}). This
internal-external validity gap may help to explain growing interest in
the use of large electronic health records (EHR) databases to inform
clinical research (\protect\hyperlink{ref-rwe_nam}{Galson and Simon
2016}).

Issues of relevance and external validity apply to an even greater
degree when synthesizing evidence across several studies. Many of these
challenges relate to possible sources of heterogeneity between the RCTs
included in a given meta-analysis. Considering only one such possible
difference, that of heterogeneity in the participant population
underlying each RCT, it is difficult to imagine how an estimated
treatment effect averaged over such populations---as is done in standard
meta-analyses---would apply to a given clinician's or health system's
population of interest. Moreover, other relevant sources of
heterogeneity remain, including treatment modalities and methods of
evaluating primary outcomes
(\protect\hyperlink{ref-jama_meta_analysis}{Berlin and Golub 2014}).

The standard approach to modeling such between-study heterogeneity is
random-effects meta-analysis, in which the effects or outcomes of each
study are conceived as random draws from some distribution, typically a
normal distribution (\protect\hyperlink{ref-re_meta}{Higgins, Thompson,
and Spiegelhalter 2009}). Heterogeneity between trials is therefore
viewed as part of the total variance in trial outcomes: one component
arising from sampling variation within a trial and another reflecting
systematic (though still i.i.d.) variation between the trials.

However, estimates from random-effects meta-analyses still fail to
explicitly account for differences between trial participants and a
target population of interest to clinicians and policy makers. Recent
work in causally interpretable meta-analysis has yielded methods for
making population-specific inferences using data from multiple RCTs
(Dahabreh et al. (\protect\hyperlink{ref-dahabreh_biometrics}{2022}),
(\protect\hyperlink{ref-dahabreh_epi}{2020})). Roughly, this work uses a
representative sample from the target population of interest to
transport treatment effects estimated in the RCT populations to the
target population.

A key assumption in this approach is that two individuals from the same
target population would have identical average responses to treatment
regardless of which clinical trial in the meta-analysis they may have
participated in. Moreoever, the treatment response observed for an
individual in the target population had they participated in any one of
the trials is assumed analogous to the treatment response expected in
the population more generally. Systematic heterogeneity in trial conduct
and the possibility of trial participation effects threaten both of
these assumptions (\protect\hyperlink{ref-dahabreh_adherence}{Dahabreh,
Robertson, and Hernán 2022}). While some of this heterogeneity might be
captured by covariate differences, other site-specific mechanisms may be
unrelated to participant characteristics.

\hypertarget{examples-of-between-study-heterogeneity}{%
\subsection{Examples of Between-Study
Heterogeneity}\label{examples-of-between-study-heterogeneity}}

Defining such heterogeneity more concretely, consider a scenario where
each study in the meta-analysis applied a somewhat different version of
treatment. In the context of studies designed to ameliorate lower back
pain, for example, providers in one study may apply a slightly different
form of spinal manipulation than providers in another study. Such
differences could be pre-defined in each study's protocol, but may also
arise simply because different chiropractors implement the manipulation
in slightly different ways. Moreover, differences in how a pain outcome
is measured (e.g., the exact time at which the pain scale is assessed
relative to treatment) could also require an expansion of potential
outcome notation.

These differences in, say, treatment version, would persist even after
accounting for variation in the distribution of treatment effect
modifiers both between studies and between the trial and target
populations. In the back pain example, if two individuals with identical
covariate data enrolled in separate studies, differences in the spinal
manipulation type would persist when attempting to compare and define
their respective potential outcomes. This suggests indexing such
heterogeneity in the potential outcomes themselves, e.g., an
individual's potential outcome had they been assigned treatment \(a\)
under one provider versus another.

Broader differences in trial conduct also have the potential to induce
this heterogeneity. For example, after disruptions to biomedical
research caused by the COVID-19 pandemic, some clinical trials opted to
deliver self-administered medications to trial participants by mail,
rather than distribute them in-person
(\protect\hyperlink{ref-fda_guidance}{U.S. Food and Drug Administration
2021}). Delivery timings under these conditions may vary significantly
between different trials, e.g., trials whose participants reside in more
remote, rural locations may have slower delivery times than trials in
urban locations. For diseases like COVID-19, where timing of treatment
affects subsequent outcomes, these differences in delivery time could
induce heterogeneity in the potential outcomes from each trial.

Differences in the timing of treatments administered in-person may also
have important implications for research synthesis. A recent preprint
examined whether passively administered antibodies altered health
outcomes associated with SARS-CoV-2 infection
(\protect\hyperlink{ref-monoclonal_antibody}{Stadler et al. 2022}). They
found that earlier administration of such treatments may improve their
efficacy. Imagine two patients with identical baseline covariates
enrolling in two studies investigating the same monoclonal antibody
treatment. If the two studies differed in the timing of treatment
relative to infection, this study suggests that those individuals may
have distinct, study-specific expected potential outcomes.

Another example where differences in trial conduct may induce
heterogeneity is when provider-assessed ratings are an outcome. For
example, The Positive and Negative Syndrome Scale (PANSS) is routinely
used to assess symptom severity in patients with schizophrenia
(\protect\hyperlink{ref-panns_general}{Kay, Fiszbein, and Opler 1987}).
For each symptom evaluated in the PANSS, raters assign a score of 1
(absent) to 7 (extreme). Differences between treatment groups might then
be assessed by differences in their aggregated PANSS scores. However,
individual raters may perform these qualitative evaluations in different
ways. Analyses synthesizing outcomes from many sites employing different
raters may need to take these differences into account.

Animated by such concerns, our work aims to lay the groundwork for a
causally-interpretable meta-analysis that accounts for heterogeneity
between studies beyond that induced by differences in the distribution
of treatment effect modifiers. Our approach attempts to combine the
techniques and intuitions of traditional random-effects meta-analysis
with the causal inference-based meta-analytic framework introduced by
Dahabreh et al. (\protect\hyperlink{ref-dahabreh_biometrics}{2022}).
Concretely, this involves posing two causal questions. First, what
outcomes would we observe, on average, if a member of the target
population had participated in a particular study included in the
collection of relevant RCTs? Moreover, what would we observe if we
conducted a new RCT whose participants are drawn directly from the
target population? This latter question is of primary scientific
interest in this work, and follows directly from our novel framework. By
answering such questions, we hope to expand the toolkit available to
health policy makers and clinicians when evaluating treatment efficacy
through meta-analysis.

\hypertarget{estimands-for-capturing-between-study-heterogeneity}{%
\section{Estimands for Capturing Between-Study
Heterogeneity}\label{estimands-for-capturing-between-study-heterogeneity}}

Causally-interpretable meta-analysis analyzes outcomes and treatment
effects observed in a collection of studies through the lens of
transportability and generalizability. We assume the existence of some
target population of interest and consider what this collection of
studies can tell us about the effect of an intervention in that target
population. Suppose we have data from a collection of clinical trials
indexed by a set \(\mathcal{S}=\{1, ..., m\}\). One causal question of
interest is: what treatment effect can we expect if a member of the
target population had participated in study \(s\in \mathcal{S}\)? As we
will describe more precisely later, we can roughly describe our approach
as decomposing such an average treatment effect \(\tau_s\) into the sum
of an overall grand mean \(\tau\) and a deviation \(\delta_s\) specific
to study \(s\): \begin{equation}
    \tau_s = \tau + \delta_s. \label{eq: caricature}
\end{equation} In such a formulation, the grand mean \(\tau\) gives the
treatment effect in the target population averaged over the collection
of studies while the \(\delta_s\) reflects between-study heterogeneity
in the underlying treatment effect.

Our ultimate goal is to precisely characterize, estimate, and perform
inference on \(\tau_0\), which is the expected treatment effect if a new
trial were conducted in the target population. Such a treatment effect
comes about via a new draw \(\delta_0\), a new shift about \(\tau\).

The decomposition in \eqref{eq: caricature} is structurally similar to
parameters studied in random-effects meta-analysis, wherein
study-specific treatment effects are assumed to be i.i.d. draws from a
distribution, typically normal. In this work, we define and estimate
parameters that maintain the familiar form of \eqref{eq: caricature}
while admitting a precise causal interpretation. This is accomplished by
incorporating the study-specific heterogeneity reflected in \(\delta_s\)
within the potential outcomes framework. Specifically, we incorporate
the variation represented by \(\delta_s\) as an additional argument to
the standard potential outcomes notation.

\hypertarget{the-data}{%
\subsection{The Data}\label{the-data}}

Adding notation to the example outlined above, suppose we have IPD from
a collection of \(m\) trials, all of which examined the same set of
treatments. Again, we index these trials with the set
\(\mathcal{S}=\{1, ..., m\}\) and treatments with the set
\(\mathcal{A}\). For each participant \(i\) in a given trial
\(s\in \mathcal{S}\) with \(n_s\) participants total, we have outcome
data \(Y_i\), baseline covariate information \(X_i\) and treatment
assignment \(A_i\), where \(A_i \in \mathcal{A}\). We also assume to
have baseline covariate information from a random sample of individuals
in a target population of interest; we do not require treatment or
outcome information in this sample. As in Dahabreh et al.
(\protect\hyperlink{ref-dahabreh_biometrics}{2022}), we let \(S=0\) for
individuals in the target population and introduce another variable
\(R\) which takes value 1 for participants in the collection of trials
and 0 for members of the target population. Thus, the observed data for
each individual in the full dataset---that is, combined data from both
the collection of trials and the target population---is of the form
\begin{equation*}
    (R_iY_i, R_iA_i, X_i, S_i).
\end{equation*} \(R_iY_i\) and \(R_iA_i\) evaluate to zero for those in
the target population, indicating that such data are unavailable. The
total sample size including both the sample from the target population
and the collection of trials is \(n=n_0+n_1+\cdots+n_m\).

\hypertarget{rationale-for-new-causal-quantities}{%
\subsection{Rationale for New Causal
Quantities}\label{rationale-for-new-causal-quantities}}

In introducing causally-interpretable meta-analysis, Dahabreh et al.
(\protect\hyperlink{ref-dahabreh_biometrics}{2022}) define potential
outcomes \(Y(a)\) that depend on the assigned treatment
\(a\in \mathcal{A}\). Different causal estimands describe the
distribution of these potential outcomes within various populations of
interest. For instance, \(E[Y(a)|R=0]\) gives the expected potential
outcome under treatment \(a\) in the target population. To identify
quantities like \(E[Y(a)|R=0]\), Dahabreh et al.~assume exchangeability
in potential outcomes across values of \(S\), conditional on baseline
covariates. They also show that \(E[Y(a)|R=0]\) is identifiable under
the weaker assumption of mean exchangeability, roughly defined as the
assumption that: \begin{equation}
    E[Y(a)|X=x, S=0]=E[Y(a)|X=x, S=s] \label{eq: dahabreh_exch}
\end{equation} for \(s=1, ..., m\), along with additional conditions
ensuring that the assumption is only made across values of \(S\) where
the covariate pattern \(X=x\) has positive probability of occurring.

Intuitively, the primary risk in assuming \eqref{eq: dahabreh_exch} when
it does not, in fact, hold is that the particular idiosyncrasies of each
trial have the potential to muddle a transported treatment effect
estimated from pooled data. This is especially problematic if the trials
in the meta-analytic collection differ a great deal in sample size. In
that setting, violations of \eqref{eq: dahabreh_exch} may result in
causal estimates that are heavily weighted to the particular conditions
of the largest trial. Such a weighting scheme has no clinical meaning
and is merely an artifact of the mean exchangeability assumption.

The motivation for referencing such between-study heterogeneity stems
directly from the logic underlying standard random-effects meta-analysis
(\protect\hyperlink{ref-re_meta}{Higgins, Thompson, and Spiegelhalter
2009}). In that paradigm, a meta-analysis of a collection of trials
proceeds under the assumption that the underlying parameter of
interest---typically an average treatment effect---differs between
trials. More precisely, in the traditional random-effects model, each
study's latent true treatment effect is assumed to be randomly sampled
from a distribution of effects. The variance of this underlying
distribution is often of interest, and serves to quantify the degree of
between-study heterogeneity.

Our proposed method models the average effects arising from studies
\(s\in \mathcal{S}\) transported to the target population as varying
about a grand mean according to draws of a latent random variable. Draws
of this latent random variable fix these underlying average effects at
different values, where differences between these values reflect
between-study heterogeneity beyond that induced by covariate
differences. The key distinction between these methods and traditional
random-effects models is that heterogeneity in underlying parameter
values (e.g., heterogeneity between studies in treatment effects) stems
from variation in the potential outcomes of study participants, thereby
retaining causal interpretations absent from standard meta-analysis.

Recall that a key causal question motivating this study is: what outcome
would we expect if an individual were assigned treatment
\(a\in \mathcal{A}\) in study \(s\in \mathcal{S}\)? This question
suggests a need to distinguish potential outcomes arising from one trial
setting versus another. Using the notation of VanderWeele and Hernán
(\protect\hyperlink{ref-mult_versions}{2013}), we write the potential
outcome of a subject receiving treatment \(a\in \mathcal{A}\) under the
conditions present in study \(s\in \mathcal{S}\) as having two
arguments: one specifying the treatment assignment and another fixing
the conditions present in a given setting, e.g., treatment group \(a\)
in study \(s\): \begin{equation*}
    Y(a, k^a_s).
\end{equation*} Here, \(k^a_s\) constitutes a realization of a random
variable \(K^a_s(a)\). Because we conceptualize values for \(K^a_s(a)\)
even for individuals who did not receive treatment \(a\) in study \(s\),
its notation mimics that of a standard potential outcome, as one's
experience in a particular study could be a function of the treatment
received. We suppose that each combination of a study and treatment arm
is associated with one such random variable, the collection of which,
across studies for a given treatment \(a\), is an i.i.d. sample from
some common distribution: \begin{equation*}
    K^a_1(a), ..., K^a_m(a)\overset{i.i.d.}{\sim} F_{K^a(a)}.
\end{equation*} We do not impose a particular interpretation on
\(K^a_s(a)\) except that it reflects the conditions an individual would
have experienced had they been in treatment group \(a\) of study \(s\).
That is, two individuals with common values of this random variable can
be understood to have been assigned treatment \(a\) under equivalent
conditions. Again, we treat \(K^a_s(a)\) as a potentially counterfactual
variable which fixes the heterogeneity in setting/treatment group
combinations even for individuals who were not assigned treatment \(a\)
or did not participate in study \(s\).

We suppose these random variables are an i.i.d. sample from some common
distribution not because we think this gives the most accurate
approximation to the data generating process that induces between-study
heterogeneity. Rather, such an assumption constitutes our best guess as
to the nature of this heterogeneity in the absence of additional
information. Equipped with only outcome, baseline covariate, and
treatment arm information, residual between-study heterogeneity is as
good as i.i.d. noise from the perspective of the (meta) analyst, playing
a role similar to residual errors in a simple linear regression model.
If more information were available about the nature of between-trial
heterogeneity, this assumption could be relaxed or refined.

The impact of this heterogeneity is related to but distinct from the
effects of trial participation per se, as studied in Dahabreh et al.
(\protect\hyperlink{ref-dahabreh_scale_up_preprint}{2019}). Here, we
develop a conceptual framework for understanding the effects of
participating in one trial versus another rather than participation in
and of itself. In part, this effort can be interpreted as clarifying the
interpretation of causal estimands defined in earlier work on
causally-interpretable meta-analysis. Recognizing the impact of
between-trial heterogeneity, we study the effects of participation in
one of the trials under study, or trials similar to those under study,
rather than a more generic effect of treatment assignment in the target
population.

\hypertarget{defining-causal-estimands}{%
\subsection{Defining Causal Estimands}\label{defining-causal-estimands}}

As in Dahabreh et al.
(\protect\hyperlink{ref-dahabreh_biometrics}{2022}), our interest is in
transporting inferences from the collection of trials to a target
population. Since the realized draw of \(K^a_s(a)=k^a_s\) indexes the
heterogeneity in potential outcomes arising from trial \(s\), a causal
quantity relevant to evaluating the kind of outcomes we would observe
had members of the target population participated in trial \(s\) is
\begin{equation}
    \mu_{a, 0}(k^a_s) = E[Y(a, k^a_s)|R=0]. \label{eq: average_s_outcome}
\end{equation} This is a fixed quantity which can be interpreted as the
average potential outcome in the target population where the
heterogeneity in application of treatment \(a\) is fixed at the value
associated with trial \(s\).

We recognize that policy makers and clinicians are often most interested
in treatment \emph{effects} within their target population, rather than
mean potential outcomes alone. In this paper, we focus our attention on
mean potential outcomes for simplicity of presentation. However,
contrasts of causal quantities like \eqref{eq: average_s_outcome}, e.g.
\begin{equation*}
\tau^{s}_{a, a'} = \mu_{a, 0}(k^a_s) - \mu_{a', 0}(k^{a'}_s) = E[Y(a, k^{a}_s)|R=0] - E[Y(a', k^{a'}_s)|R=0]
\end{equation*} can be defined corresponding to average treatment
effects in the target population under the conditions of trial \(s\).
Focus on such effects also allows relaxation of mean exchangeability
assumptions on mean potential outcomes to be replaced by mean effect
exchangeability.

Returning our focus to \eqref{eq: average_s_outcome}, note that the
random variables whose \(m\) realized values give these transported
potential outcomes for each study are a function of the latent random
sample \(K^a_1(a), ..., K^a_m(a)\): \begin{equation}
    \mu_{a,0}(a, K_1^a(a)), ..., \mu_{a,0}(a, K_m^a(a)) \label{eq: random_sample}
\end{equation} As such, they also constitute a simple random sample from
some distribution.

We assume this random sample
\(\mu_{a, 0}\left(K^a_1(a)\right), ..., \mu_{a, 0}\left(K^a_m(a)\right)\)
arises from a distribution with mean \(\mu_{a, 0}\) and finite variance.
Without any further modeling assumptions, we can decompose each such
random variable in the sample as: \begin{equation}
    \mu_{a, 0}\left(K^a_s(a)\right) = \mu_{a, 0} + \Delta^a_{0, s}, \label{var_decomp}
\end{equation} where \(\Delta^a_{0, s}\) is a random variable with mean
zero and finite variance. Note that (\ref{var_decomp}) does not impose
any particular modeling assumption on
\(\mu_{a, 0}\left(K^a_s(a)\right)\); it simply labels random variation
about its expectation over \(K^a_s(a)\) as \(\Delta^a_{0, s}\).
Rearranging (\ref{var_decomp}), we have \begin{align*}
    \Delta^a_{0, s} &= \mu_{a, 0}\left(K^a_s(a)\right) - \mu_{a, 0}\\
    &= \mu_{a, 0}\left(K^a_s(a)\right) - E_{K^a(a)}\left[\mu_{a, 0}\left(K^a_s(a)\right)\right].
\end{align*} The decomposition in (\ref{var_decomp}) allows us to better
understand the role that the latent random variables \(K^a_s(a)\) play
in driving systematic differences between studies in causal quantities.
Namely, a draw of \(K^a_s(a)=k^a_s\) corresponds to a realization of
\(\Delta^a_{0, s}=\delta^a_{0, s}\) which in turn shifts the average
potential outcome under \(a\) transported from study \(s\) to the target
population.

Recalling the stylized version of our approach in Equation
(\ref{eq: caricature}), our model for the treatment effect we would
observe if a member of the target population had been assigned to
treatment \(a\) vs.~\(a'\) in study \(s\) is therefore \begin{align}
\mu_{a, 0}(k^a_s) - \mu_{a', 0}(k^{a'}_s) &= \mu_{a, 0}+\delta^a_{0, s}-\left(\mu_{a', 0}-\delta^{a'}_{0, s}\right) \nonumber \\
&= \left(\mu_{a, 0}-\mu_{a', 0}\right)+\left(\delta^a_{0, s}-\delta^{a'}_{0, s}\right). \label{eq: te_causal_quantity}
\end{align} We can conceptualize
\(\tau_{a, a', 0} = \mu_{a, 0}-\mu_{a', 0}\) as the ``grand mean
effect'' of treatment \(a\) vs.~\(a'\) in the target population and
\(\delta^a_{0, s}-\delta^{a'}_{0, s}\) as the heterogeneity of that
overall effect when transporting from the setting of study \(s\). These
two quantities stand in for \(\tau\) and \(\delta_s\), respectively in
Equation (\ref{eq: caricature}).

Besides considering the expected potential outcome if an individual from
the target population participated in one of the trials in our
collection, we might also ask: what potential outcomes would we expect
if a new trial were conducted that recruited a simple random sample from
the target population? We make an important assumption here that the
same random process which induces heterogeneity in expected potential
outcomes within the collection of trials would apply to a new trial of
the same collection of treatments.

We can represent this assumption by adding \(K^a_0(a)\) to our original
random sample: \begin{equation*}
    K^a_1(a), ..., K^a_m(a), K^a_0(a)\overset{i.i.d.}{\sim} F_{K^a(a)}
\end{equation*} and defining \(\mu_{a, 0}\left(k^a_0\right)\) and
\(\mu_{a, 0}\left(K^a_0(a)\right)\) as previously.

This framing is analogous to the idea in random-effects meta-analysis
that we can make inferences for the treatment effects of studies that
are not included in the meta-analysis
(\protect\hyperlink{ref-re_meta}{Higgins, Thompson, and Spiegelhalter
2009}). Again, our i.i.d. assumption regarding \(K^a_0(a)\) reflects our
best guess as to the random process governing between-study
heterogeneity. In the absence of additional information, we posit both
that the same sources of heterogeneity that induce systematic
differences between RCTs in the trial sample would apply to a
hypothetical RCT in the target population and that this additional
variation stems from i.i.d. draws from a common distribution. As
explained below, we use this assumption when producing prediction
intervals that contain \(\mu_{a, 0}\left(k^a_0\right)\) with some
specified probability.

There are several possible motivations for estimating the outcomes in an
unobserved trial. One is as an input to planning and preparation for new
clinical studies. Clinicians may desire to estimate treatment effects
they might expect in a planned study with participants similar to
members of the target population. Alternatively, transported estimates
from a clinical trial might serve as a kind of bound for effects we are
likely to observe in practice. Estimating outcomes in an unobserved
trial also produces such a bound, albeit one not overly influenced by
the idiosyncrasies present in any particular study in the collection of
trials.

\hypertarget{identification-and-estimation-of-causal-quantities}{%
\section{Identification and Estimation of Causal
Quantities}\label{identification-and-estimation-of-causal-quantities}}

\hypertarget{identification}{%
\subsection{Identification}\label{identification}}

We first consider identification of \(\mu_{a, 0}(k^a_s)\), which gives
the expected outcome that would have been observed if a member of the
target population had been assigned treatment \(a\) in study \(s\). To
express this as a functional of the observed data, we make the following
identifying assumptions, many of which are similar to those in Dahabreh
et al. (\protect\hyperlink{ref-dahabreh_biometrics}{2022}) and
VanderWeele and Hernán (\protect\hyperlink{ref-mult_versions}{2013}):

\begin{enumerate}
    \item \textbf{Exchangeability in mean between trials:}
    \begin{equation*}
        E[Y(a, k^a_s)|X=x, S=s_1] = E[Y(a, k^a_s)|X=x, S=s_2]
    \end{equation*}
    for all $s_1, s_2\in \{0, 1, ..., m\}$ and $x\in \mathcal{X}$ such that $f(x, S=s_1)\neq 0$ and $f(x, S=s_2)\neq 0.$ Note here     that the heterogeneity in trial $s_1$ and $s_2$ is fixed at $k^a_s$ in both settings.
    \item \textbf{Exchangeability over treatment groups within a trial:}
    \begin{equation*}
        Y(a, k^a_s)\perp\!\!\!\perp A|(X, S=s)
    \end{equation*}
    for all $a\in \mathcal{A}$, $k^a_s\in \mathcal{K}^a$, and $s\in \mathcal{S}$. 
    \item \textbf{Consistency:} If $A_i=a$ and $K^a_i(a)=k^a$ for individual $i$, then $Y_i(a, k^a)=Y_i$.
    \item \textbf{Positivity:} For all $s\in \mathcal{S}$, if $f(x, R = 0) \neq 0$, then $P(S = s|X = x) > 0$.
    \item \textbf{Distribution of $K^a(a)$:} For $s=1, ..., m$ and $a\in\mathcal{A}$ each random variable $K^a_s(a)$ consitutes an i.i.d. draw from a distribution $F_{K^a(a)}$.
    \item \textbf{Constancy of $K^a(a)$ by treatment group/study:} Within each treatment group $a$ and study $s$, the value of $K^a(a)$ for each participant is fixed at the realized value of $K^a_s(a) = k^a_s$. That is, if $A=a$ and $S=s$, then
    \begin{equation*}
        K^a_i(a)=K^a_s(a)=k^a_s
    \end{equation*}
for $i\in 1, ..., n_s$.
\end{enumerate}

Assumptions 3 and 5 involve a slight abuse of notation, wherein we
define a random variable \(K^a_i(a)\) giving the version of treatment
for individual \(i\). The key point here is that such a random variable
takes on the same realized value of \(K^a_s(a)\) for all participants in
trial \(s\). Alternatively, one might assume a hierarchical model where
each individual in treatment arm \(a\) and study \(s\) has an associated
random variable \(\left(K^a_s(a)\right)_{i}\), \(i=1, ..., n_s\),
centered at a the realized value of \(K^a_s(a)=k^a_s\). In this paper,
we focus on trial/treatment group-wide heterogeneity to greatly simplify
the mathematical presentation. However, a hierarchical model of the kind
proposed above can--under certain assumptions--lead to the same
identification results presented here.

Under these assumptions, we identify \(\mu_{a, 0}(k^a_s)\) as

\begin{equation}
\psi_{s, 0}(a) = E[E[Y|X, S=s, A=a]|R=0]. \label{eq: id_one_trial}
\end{equation}

That is, we average a regression function relating outcomes under
treatment \(a\) in study \(s\) to covariates \(X\) over the distribution
of such covariates in the observed target population. (A full proof of
this result is given in the Appendix.) This observed data functional is
analogous to that identified in Equation (6) of Theorem 1 in Dahabreh et
al. (\protect\hyperlink{ref-dahabreh_biometrics}{2022}), with \(R=1\) in
their case replaced by \(S=s\) in ours.

\hypertarget{estimation}{%
\subsection{Estimation}\label{estimation}}

Letting \(g^s_{a}(X)=E[Y|X, S=s, A=a]\), we could apply an outcome
model/standardization approach that averages an estimate
\(\hat{g}^s_{a}(X)\) of \(E[Y|X, S=s, A=a]\) over the distribution of
covariates in the target population: \begin{equation*}
\hat{\psi}_{s, 0}(a)=\left\{\sum_{i=1}^{n}I(S_i=0)\right\}^{-1}\sum_{i=1}^{n}I(S_i=0)\left\{\hat{g}^s_{a}(X_i)\right\} \label{eq: om_estimator}
\end{equation*}

Another option involves inverse probability weighting, in which the
outcomes observed among participants in study \(s\) are weighted by the
similarity of each participant to members of the target population. The
estimator applying this approach is given by \begin{equation*}
\hat{\psi}^{ipw}_{s, 0}(a)=\left\{\sum_{i=1}^{n}I(S_i=0)\right\}^{-1}\sum_{i=1}^n \left(\frac{I(A_i=a)}{\hat{e}^s_a(X_i)}\right)I(S_i=s)\frac{\hat{p}_0(X_i)}{\hat{p}_s(X_i)}Y_i
\end{equation*} where \(\hat{p}_0(X_i)\) estimates \(P(R=0|X_i)\),
\(\hat{e}_a(X_i)\) estimates \(P(A_i=a|S=s, X_i)\), and
\(\hat{p}_s(X_i)\) estimates \(P(S=s|X_i)\). Identification of this
result, proceeding from \eqref{eq: id_one_trial} is given in the
appendix.

A final option is the so-called augmented inverse probability weighting
(AIPW) estimator, which combines the two approaches introduced above:

\begin{equation*}
\hat{\psi}^{aipw}_{s, 0}(a)=\left\{\sum_{i=1}^{n}I(S_i=0)\right\}^{-1}\sum_{i=1}^n\left[I(S_i=0)\left\{\hat{g}^s_{a}(X_i)\right\}+\left(\frac{I(A_i=a)}{\hat{e}^s_a(X_i)}\right)I(S_i=s)\frac{\hat{p}(X_i)}{1-\hat{p}(X_i)}(\hat{g}^s_{a}(X_i)-Y_i)\right]
\end{equation*}

This approach adds a correction to the outcome model approach via
inverse probability weighting over the residuals of the outcome model
for arm \(a\) of study \(s\).

Our second causal quantity of interest is \(E[Y(a, k^a_0)|R=0]\), where
\(k^a_0\) is the realized value of the random variable \(K^a_0(a)\)
described in the previous section. In the absence of additional
information concerning the distribution \(F_{K^a(a)}\), our estimation
strategy for \(E[Y(a, k^a_0)|R=0]\) relies on the following
approximation, the full details of which are included in the Appendix:
\begin{align*}
E[Y(a, K^a_0(a))|R=0] &= \sum_{k^a\in \mathcal{K}^a}E\left\{E[Y(a, K^a_0(a))|R=0, K^a_0(a)=k^a]|R=0\right\}P(K^a_0(a)=k^a)\\
    &\approx \sum_{k^a\in\mathcal{K}^a}E\left\{E[Y(a, K^a_0(a))|R=0, K^a_0(a)=k^a]|R=0\right\}\left\{\frac{1}{m}\sum_{s=1}^m \mathbbm{1}(k^a=k^a_s)\right\}\\
    &= \frac{1}{m}\sum_{s=1}^m E[Y(a, k^a_s)|R=0]\\
    &= \frac{1}{m}\sum_{s=1}^m E[E[Y|X, S=s, A=a]|R=0] \hspace{3.4cm} \text{by \eqref{eq: id_one_trial}}\\
    &= \frac{1}{m}\sum_{s=1}^m \psi_{s, 0}(a).
\end{align*}

Again, taking contrasts of the above quantity applied to different
treatments implies identification of treatment effects of interest,
e.g., causal quantities of the form given in
(\ref{eq: te_causal_quantity}). We can then estimate
\(\frac{1}{m}\sum_{s=1}^m \psi_{s, 0}(a)\) using any of the methods
discussed above for estimating each \(\psi_{s, 0}(a)\) individually. For
instance, we might estimate this quantity using \begin{equation*}
\frac{1}{m}\sum_{s=1}^m \hat{\psi}_{s, 0}(a) = \frac{1}{m}\sum_{s=1}^m \left(\left\{\sum_{i=1}^{n}I(S_i=0)\right\}^{-1}\sum_{i=1}^{n}\left\{\hat{g}^s_{a}(X_i)\right\}\right).
\end{equation*}

\hypertarget{estimation-of-between-study-variability}{%
\subsection{Estimation of Between-Study
Variability}\label{estimation-of-between-study-variability}}

A central parameter of interest in traditional random-effects
meta-analysis is the between-study variance, which describes the
variability of the effects underlying each study. Recalling the
decomposition given in (\ref{var_decomp}), the analogous parameter in
our work is the variance of \(\Delta^a_{0, s}\), which specifies the
variability between studies in the transported mean potential outcomes.
We propose an estimate of this variance inspired by the method of
moments approach derived by Rao et al.
(\protect\hyperlink{ref-tau_paper}{1981}) and subsequently applied by
DerSimonian and Laird (\protect\hyperlink{ref-dersimonian}{1986}) in
their seminal work on random-effects meta-analysis. The key distinction
between our estimate and the traditional random-effects estimate is
that, in practice, the estimates for
\(\mu_{a, 0}(k^a_1), ..., \mu_{a, 0}(k^a_m)\) are correlated due to
their dependence on the same sample from the target population. This
correlation complicates the derivation and form of the resulting
estimator.

In the derivation below, we consider estimating between-study
variability in a highly general setting; the only relationship to our
causal framework is that of correlation between study-specific
estimates. Operating in a simplified setting, suppose we have \(m\)
study-specific means \(\mu_1, ..., \mu_m\) drawn from a distribution
\(F_\mu\) with mean \(\mu\) and variance \(\gamma^2\). The analogy to
our setting is obtained by letting \(\mu_s=\mu_{a, 0}(k^a_s)\). We
estimate these means with \(\hat{\mu}_1, ..., \hat{\mu}_m\) each of
which is individually unbiased for its respective study-specific mean.
The analogous quantities for us are
\(\hat{\mu}_s=\hat{\psi}_{s, 0}(a)\). We estimate the grand mean \(\mu\)
using a weighted sum of the study-specific estimates
\(\hat{\mu}=\sum_{s=1}^m w_s\hat{\mu}_s\), where \(\sum_{s=1}^m w_s=1\).
(For instance, we might have \(w_s=\frac{1}{m}\) when taking a simple
average.) Also, let
\(\text{Var}\left(\hat{\mu}_s\right|S=s) = \sigma^2_s\) denote the
sampling variance of each study-specific estimator and
\(\text{Var}\left(\hat{\mu}_s\right) = \sigma^2_s + \gamma^2\), which
reflects both within- and between-study variance. Letting
\(Q = \sum_{s=1}^m \left(\hat{\mu}_s-\hat{\mu}\right)^{2}\) and
\(C_s = -2\sum_{i\neq s}w_i(1-w_s)\sigma_{is}+\sum_{i\neq s}\left[\sum_{j\neq i, j\neq s}w_iw_j\sigma_{ij}\right]\)
we can show that \begin{equation*}
    E[Q]=m\sum_{s=1}^m w_s^2(\sigma^2_s+\gamma^2)+\sum_{s=1}^m \left[(1-2w_s)(\sigma^2_s+\gamma^2)+C_s\right].
\end{equation*} A moment-based estimator for \(\gamma^2\) is then given
by the value \(\Tilde{\gamma}^2\) that satisfies \begin{equation*}
    \sum_{s=1}^m \left(\hat{\mu}_s-\hat{\mu}\right)^2 = m\sum_{s=1}^m w_s^2(\sigma^2_s+\Tilde{\gamma}^2)+\sum_{s=1}^m \left[(1-2w_s)(\sigma^2_s+\Tilde{\gamma}^2)+C_s\right].
\end{equation*} Solving for such a \(\Tilde{\gamma}^2\), we obtain
\begin{equation*}
    \Tilde{\gamma}^2 = \frac{\sum_{s=1}^m \left(\hat{\mu}_s-\hat{\mu}\right)^2-\sum_{s=1}^m \sigma^2_s\left\{mw_s^2+(1-2w_s)\right\}-\sum_{s=1}^m C_s}{\sum_{s=1}^m\left\{mw_s^2+(1-2w_s)\right\}}.
\end{equation*} Our final estimate for the between-study heterogeneity
would then be \begin{equation}
    \hat{\gamma}^2 = \max(0, \Tilde{\gamma}^2). \label{eq: our_gamma_squared}
\end{equation} A full derivation of the above estimator is given in the
Appendix. The key distinction between our estimator and the traditional
random-effects estimator introduced by Rao et al.
(\protect\hyperlink{ref-tau_paper}{1981}) is the inclusion of
\(-\sum_{s=1}^m C_s\) in the numerator of \(\hat{\gamma}^2\). This has
the effect of ``correcting'' the traditional estimator to account for
the nonzero correlation between study-specific estimates. Beyond our
setting, the estimator can be generally applied to any analysis which
takes a weighted average of correlated quantities to estimate some
underlying grand mean; note, though, that it is subject to bias when
individual estimates are themselves biased.

\hypertarget{inference-for-outcomes-in-a-new-trial}{%
\subsection{Inference for Outcomes in a New
Trial}\label{inference-for-outcomes-in-a-new-trial}}

Inference for the outcomes we would expect if members of the trial
population participated in an \emph{observed} trial, e.g.,
\(\mu_{a, 0}(k^a_s)\), \(s\in \mathcal{S}\) can proceed in a variety of
ways, including simple bootstrap resampling on data from the target
population and study \(s\) or via asymptotic approximations to the
sampling distribution of \(\hat{\psi}_{s, 0}(a)\). Here, we focus on
inference for \(\mu_{a, 0}(K^a_0(a))\). We keep \(K^a_0(a)\) as a random
variable in this estimand since it refers to outcomes in a trial we have
yet to observe. As such, a simple bootstrap applied to all observed
trials and the target population would fail to capture the excess
variability induced by conducting a new trial. That is, our prediction
intervals aim to capture both the sampling variability of our estimator
\(\frac{1}{m}\sum_{s=1}^m \hat{\psi}_{s, 0}(a)\) as reflected in the
observed data and additional uncertainty about the unobserved trial. We
can make progress under the assumption that the unobserved trial is
subject to similar heterogeneity as that between the observed trials.
Here, we propose three methods for constructing these prediction
intervals, and later evaluate them in a simulation experiment.

\hypertarget{inference-based-on-hatgamma2}{%
\subsubsection{\texorpdfstring{Inference Based on
\(\hat{\gamma}^2\)}{Inference Based on \textbackslash hat\{\textbackslash gamma\}\^{}2}}\label{inference-based-on-hatgamma2}}

Above, we derived an estimator \(\hat{\gamma}^2\), which we generically
interpret as an estimate of between-study heterogeneity beyond that
induced by measured covariates. When we replace \(\hat{\mu}_s\) with
\(\hat{\psi}_{s, 0}(a)\) and \(\hat{\mu}\) with
\(\frac{1}{m}\sum_{s=1}^m \hat{\psi}_{s, 0}(a)\), \(\hat{\gamma}^2\)
then estimates
\(\text{Var}\left(\mu_{a, 0}\left(K^a_s(a)\right)\right)\). That is,
within our framework, \(\hat{\gamma}^2\) quantifies the variability in
expected transported potential outcomes from one trial to another, where
the variability is driven by setting-specific heterogeneity. To
construct a prediction interval centered at
\(\frac{1}{m}\sum_{s=1}^m \hat{\psi}_{s, 0}(a)\), we employ the same
form of the interval considered in Higgins et al.
(\protect\hyperlink{ref-re_meta}{2009}) for inference on ``the effect in
an unspecified study'': \begin{equation}
\frac{1}{m}\sum_{s=1}^m \hat{\psi}_{s, 0}(a)\pm t^\alpha_{m-2}\sqrt{\hat{\gamma}^2 + \widehat{\text{Var}}\left(\frac{1}{m}\sum_{s=1}^m \hat{\psi}_{s, 0}(a)\right)} \label{eq: gamma_interval},
\end{equation} where
\(\widehat{\text{Var}}\left(\frac{1}{m}\sum_{s=1}^m \hat{\psi}_{s, 0}(a)\right)\)
is obtained using a simple bootstrap on all of the observed studies and
the target population, and \(t^\alpha_{m-2}\) is the \((1-\alpha)\)th
quantile of the \(t\) distribution with \(m-2\) degrees of freedom.

\hypertarget{inference-based-on-the-bootstrap}{%
\subsubsection{Inference Based on the
Bootstrap}\label{inference-based-on-the-bootstrap}}

One way to approximate the difference between our observed estimates and
that of an unobserved trial is to compute
\(\frac{1}{m-1}\sum_{s=1}^{m-1} \hat{\psi}_{s, 0}(a)\) using data from a
subset of \(m-1\) trials and compare this result with that of the
transported potential outcome from the \(m\)th, left-out trial (as
repeated for all trials). This procedure aims to capture variability
related to between-trial heterogeneity. To additionally capture sampling
variability in the observed data, we couple the leave-one-out approach
with application of the simple or wild bootstrap. Both such bootstraps
are described below. The result of both procedures is to construct an
estimate for the distribution of \(\mu_{a, 0}\left(K^a_s(a)\right)\).
For ease of notation in the following, we let
\(\hat{\mu}_0=\frac{1}{m}\sum_{s=1}^m \hat{\psi}_{s, 0}(a)\) and
\(\hat{\mu}_s=\hat{\psi}_{s, 0}(a)\).

Let \(D_s\) denote all of the data available in our original sample from
trial \(s=1, ..., m\). Let \(X\) denote the covariate data available
from the target population. For \(b=1, ..., B\),

\begin{enumerate}
\itemsep-0.25em
    \item Randomly choose one of $m$ studies to treat as the ``unobserved'' trial on which we're trying to make a prediction. Denote this choice $s_b\in \{1, ..., m\}$.
    \item Draw a simple bootstrap sample $D^*_{s_b}$ from $D_{s_b}$ and $X^*_1$ from $X$ and estimate $\hat{\mu}^*_{s_b}$ using these two datasets.
    \item Draw simple bootstrap samples $\{D^*_s: s\neq s_b\}$ from the other trial data. Draw another bootstrap sample $X^*_2$ from $X$. With these samples, compute $\mu^*=\frac{1}{m-1}\sum_{s\neq s_b}\hat{\mu}^*_s$.
    \item Compute an estimated residual $\hat{\delta}^*_b = \mu^*-\hat{\mu}^*_{s_b}$ that quantifies our prediction error
    \item Finally, construct another estimate $\hat{\mu}^*$ using a new set of bootstrap samples for all of the original data, including the trial that we left out when computing the prediction error. Let $\hat{\mu}^{pred}_b = \hat{\mu}^*-\hat{\delta}^*_b$.
\end{enumerate}

We also consider constructing prediction intervals using the wild
bootstrap based on the influence function of the estimator
\(\hat{\mu}_s\). This procedure follows an outline similar to the simple
bootstrap above, except that the approach of Matsouaka et al.
(\protect\hyperlink{ref-if_wild_bootstrap}{2022}) replaces sampling the
data with replacement. Note that the exact form of the influence
function depends on the estimator. Because we use the outcome model in
the simulation study described below, we use its influence function in
implementing the wild bootstrap. See, e.g., Tsiatis
(\protect\hyperlink{ref-tsiatis_semiparametric}{2006}) for additional
detail regarding the definition and derivation of influence functions.

\hypertarget{simulation-study}{%
\section{Simulation Study}\label{simulation-study}}

To evaluate the performance of the three approaches outlined above, we
conduct a simulation study as follows.

\hypertarget{data-generating-procedure}{%
\subsection{Data Generating Procedure}\label{data-generating-procedure}}

\textit{Covariates}: We assume we have three covariates \(X_1\),
\(X_2\), \(X_3\) for each of the \(m\) trials and the sample from the
target population. Each trial has 100 participants total, split into two
treatment groups, and the target population sample is of size 1000. The
covariates in each setting \(s=0, 1..., m\) are generated as follows:

\begin{enumerate}
    \item Determine the mean covariate vector $\pmb{\mu}_s\in \mathbb{R}^3$ in study $s$ by generating $m$ equally spaced values in the interval $[0, 1.5]$ and setting $\pmb{\mu}_s$ equal to the $s$th such value (repeating the value three times for each entry in $\pmb{\mu}_s$). Thus, if $m=3$, then $\pmb{\mu}_1=(0, 0, 0)^T$, $\pmb{\mu}_2=(0.75, 0.75, 0.75)^T$, and $\pmb{\mu}_3=(1.5, 1.5, 1.5)^T$. In every scenario, the mean covariate vector in the target population is $\pmb{\mu}_0=(1, 1, 1)^T$. 
    \item For each of $n_s$ participants in study $s$, draw a vector $(\mathbf{X}_s)_i\sim \mathcal{N}(\pmb{\mu}_s, \Sigma)$, where
    \begin{equation*}
        \Sigma = \begin{pmatrix} 1 & 0.5 & 0.5\\
        0.5 & 1 & 0.5\\
        0.5 & 0.5 & 1\end{pmatrix}.
    \end{equation*}
    Note that this matrix is identical across studies and the target population. 
\end{enumerate}

\textit{Potential Outcomes}: Equipped with these covariate values, we
generate each participant's potential outcome under treatment \(a\) in
study \(s\) as a linear combination of
\((X_{s, 1}, X_{s, 2}, X_{s, 3})_i^T\). That is, \begin{equation}
    \left(Y(a, k^a_s)|S=s, \mathbf{X}=\mathbf{x}\right)_i=\beta_0+\beta_1x_{s, 1}+\beta_2x_{s, 2}+\beta_3x_{s, 3}+\delta^a_s+\epsilon, \label{eq: potential_outcome_sim}
\end{equation} where \(\beta_0=\beta_1=\beta_2=\beta_3=0.5\) and
\(\epsilon\sim N(0, 1)\). The value \(\delta^a_s\) is the realized value
of random variation in potential outcomes that results from applying
treatment \(a\) in study \(s\); it is fixed across all participants in
study \(s\). The distribution of this variation is distinct in different
simulation scenarios, as detailed below. Under the specification in
(\ref{eq: potential_outcome_sim}), the true value for the estimand of
interest is \begin{align}
    E[Y(a, k^a_0)|R=0]=E[E[Y(a, k^a_0)|S=0, \mathbf{X}]] &= 0.5+0.5E[X_{0, 1}]+0.5E[X_{0, 2}]+0.5E[X_{0, 3}]+\delta^a_0+E[\epsilon]\nonumber \\
    &=0.5+0.5(1)+0.5(1)+0.5(1)+\delta^a_0+0\nonumber\\
    &= 2+ \delta^a_0.\label{eq: true_target_outcome}
\end{align} This is the true expected value under treatment \(a\) if
members of the target population participated in an ``unobserved''
trial. Treatment assignment in each observed trial to \(a\) or placebo
proceeds under 1:1 randomization.

\textit{Estimation}: Using notation defined earlier, we estimate the
average observed outcome under treatment \(a\) in study \(s\) (denoted
\(g^s_a(x)=E[Y|X=x, S=s, A=a]\)) using a correctly specified outcome
model \begin{equation*}
    \hat{g}^s_a(x)=\mathbf{x}^T\hat{\pmb{\beta}}_s,
\end{equation*} where \(\mathbf{x}^T=(1, x_1, x_2, x_3)^T\) and
\(\pmb{\beta}_s=(\beta_{s, 0}, \beta_{s, 1}, \beta_{s, 2}, \beta_{s, 3})^T\).
The expected value of this outcome averaged over the distribution of
covariates in the target population (denoted
\(\psi_{s, 0}(a)=E[E[Y|X, S=s, A=a]|R=0]\)) is estimated as
\begin{equation*}
    \hat{\psi}_{s, 0}(a)=\frac{1}{n_0}\sum_{i: S_i=0}\hat{g}^s_a(X_i).
\end{equation*} As described above, the average of the
\(\psi_{s, 0}(a)\)'s over the \(m\) studies is our best guess of the
target estimand \(\mu_{a, 0}(k^a_0)\): \begin{equation*}
    \hat{\mu}_{a, 0}(k^a_0)=\frac{1}{m}\sum_{s=1}^m \hat{\psi}_{s, 0}(a).
\end{equation*} We then construct a prediction interval centered at
\(\hat{\mu}_{a, 0}\) which aims to contain
\(E[Y(a, k^a_0)|R=0]=2+\delta^a_0\) with some pre-specified probability,
using the three methods described above.

\hypertarget{simulation-scenarios}{%
\subsection{Simulation Scenarios}\label{simulation-scenarios}}

\noindent The above data generating process is highly simplified. It
assumes the true outcome model is identical across scenarios and that we
specify this model correctly. Future work will complicate this setup to
investigate other issues, e.g, model misspecification. For now, this
simple scenario focuses our attention on two main parameters, which we
vary across simulations:

\begin{enumerate}
    \item \textit{The number of trials in the meta-analysis study dataset}. This takes values across $m= $5, 10, 15, 30, and 50 studies. While we rarely expect to have 50 studies in a meta-analysis, the goal of that scenario is to investigate whether a given method over/under covers as we collect data from more and more studies.
    \item \textit{The distribution of the setting-specific variation}. Settings include:, $\text{Unif}[-2, 2]$, $N(0, 1)$, $\text{Exponential}(1)-1$, and $\text{Pareto}(1, 3)$.
\end{enumerate}

The settings above define \(5\times 4=20\) distinct simulation
scenarios. We evaluate each scenario with 1000 artificial datasets
simulated from the data generating process described above. Where
applicable, any application of the bootstrap in our estimation procedure
includes 1000 replications. For a given method of constructing the
prediction interval of interest, we estimate the coverage probability as
the proportion of prediction intervals across those 1000 iterations that
contain the true parameter value.

\hypertarget{results}{%
\subsection{Results}\label{results}}

Figure 1 displays estimates of coverage for prediction intervals
constructed according to the quantiles of the bootstrap estimates. The
second set of plots does so using using an approximation to the normal
distribution,, i.e., the mean of the bootstrapped estimates \(\pm1.96\)
times the standard deviation of the bootstrapped estimates. In the
figure, Method of Moments refers to intervals constructed using our
estimate of \(\hat{\gamma}^2\), as in \eqref{eq: gamma_interval}. (Note
that the Method of Moments results are the same in both figures because
they do not depend on such choices.)

In general, the method of moments intervals based on the corrected
estimator of between-study variance perform better than bootstrap-based
alternatives across different distributions of residual heterogeneity.
The normal-based approximations for the bootstrap approaches also
outperform prediction intervals based on empirical quantiles, though the
normal-based approximations are still outperformed by the method of
moments estimator. Although simulations assuming a very large number of
MA studies are helpful for understanding aspects of our approaches'
asymptotic behavior, we recognize that 15 or more studies is larger than
the vast majority of meta-analysis datasets, especially given the need
for IPD. Thus, these simulation results suggest applying the method of
moments intervals based on the corrected estimator of between-study
variance in most circumstances when researchers are interested in
estimating effects measured in a new, unobserved trial recruited from
the target population.

Although the performance of the two bootstrap alternatives is similar,
we note that the wild bootstrap has non-negligible computational
advantages. On a standard Windows machine with an Intel Core i7-9700
Processor, the wild bootstrap approach generated prediction intervals
more than twice as fast as the simple bootstrap in a scenario with 5
studies and the sample sizes given above. The absolute times in such
cases were negligible (16 seconds vs.~6 seconds); however, with more
studies and a much larger target population sample, the wild bootstrap
becomes far more computationally attractive.

\newpage

\begin{figure}[!ht]

{\centering \includegraphics{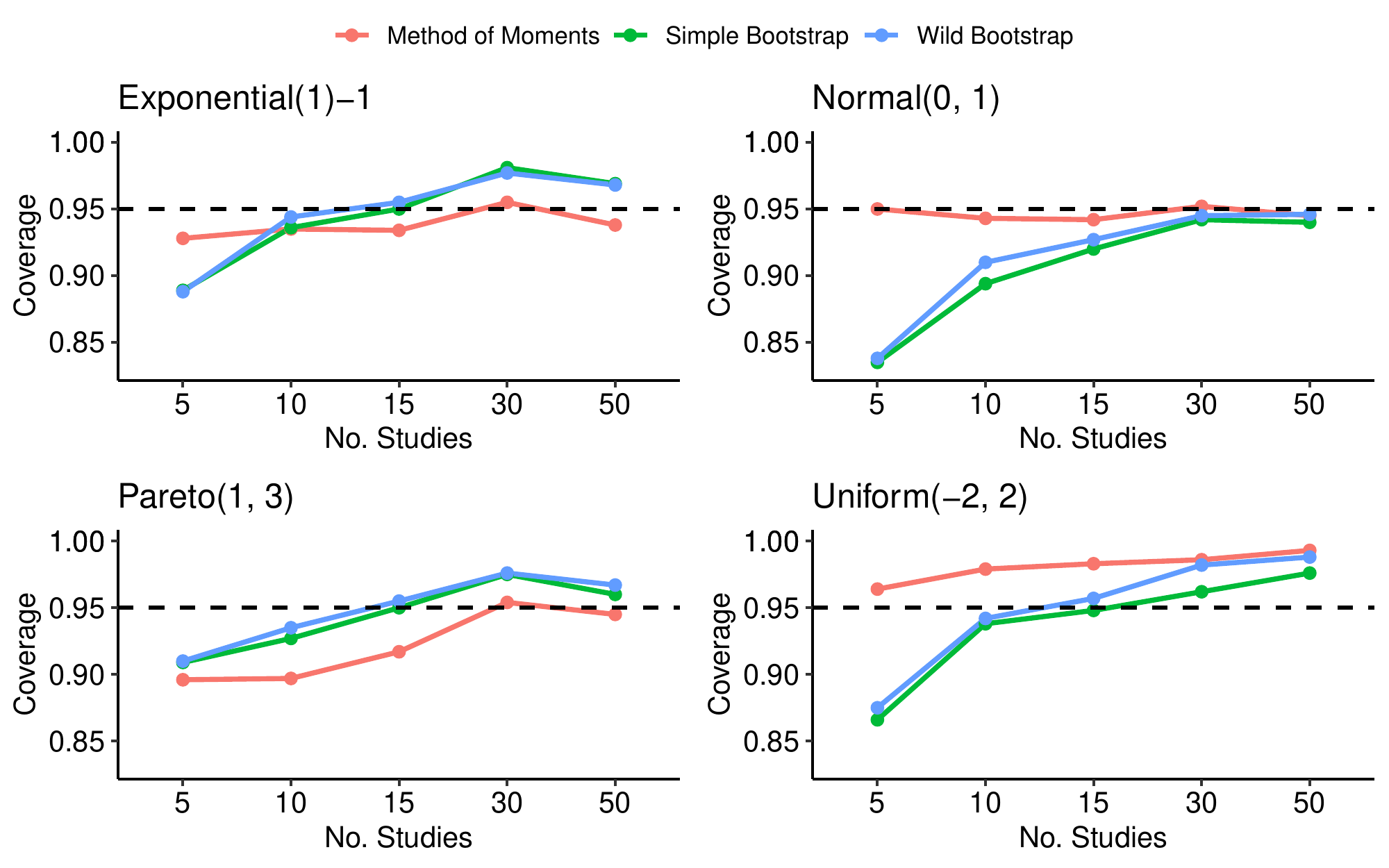} 

}

\caption{Coverage for prediction intervals constructed according to the quantiles of the bootstrap estimates. Each plot corresponds to a separate distribution for setting-specific variation.}\label{fig:quantile_based_intervals}
\end{figure}

\newpage

\begin{figure}[!ht]

{\centering \includegraphics{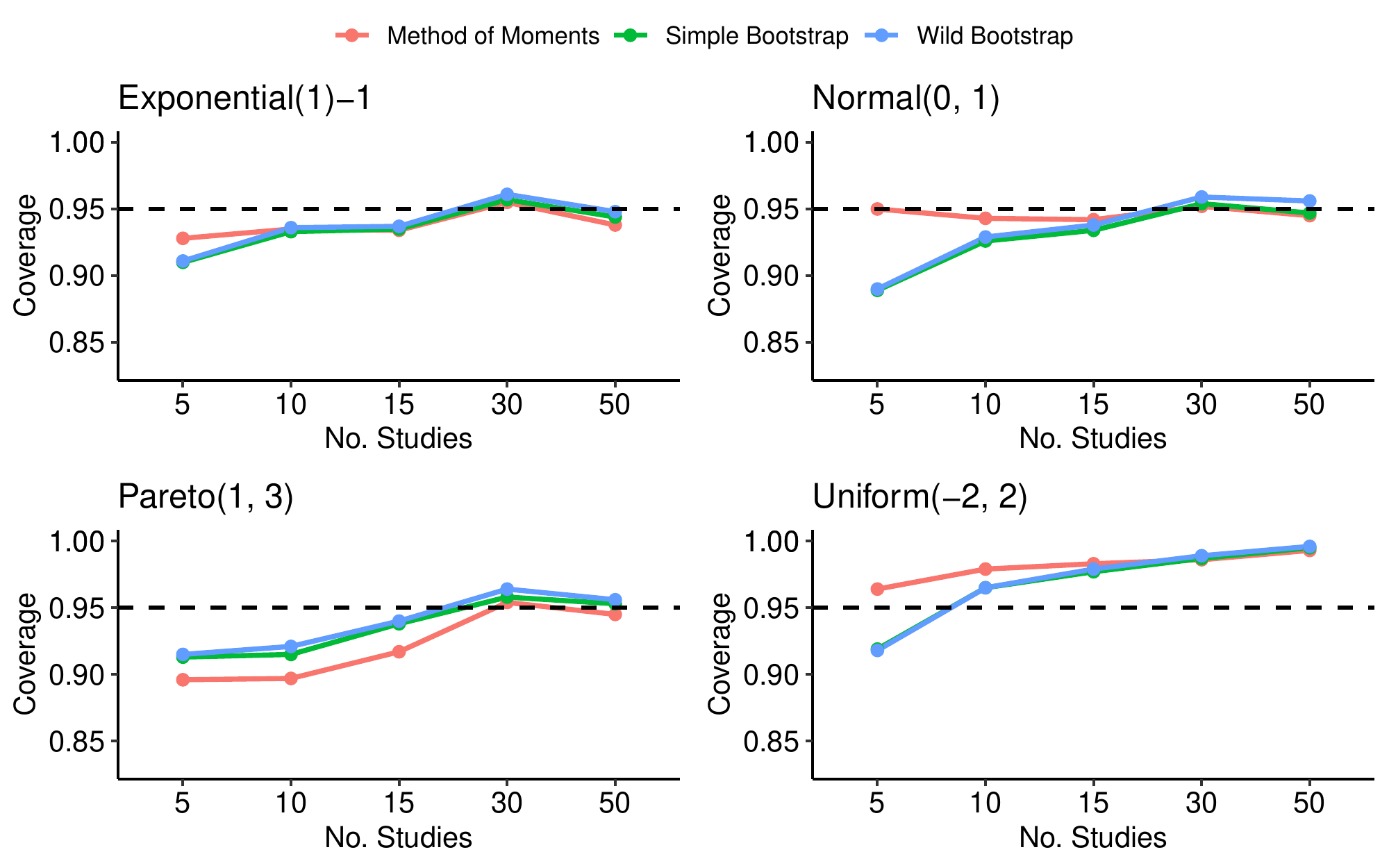} 

}

\caption{Coverage for prediction intervals constructed according to the quantiles of the normal distribution. Each plot corresponds to a separate distribution for setting-specific variation.}\label{fig:normal_based_intervals}
\end{figure}

\newpage

\hypertarget{discussion}{%
\section{Discussion}\label{discussion}}

In this work, we extend recent developments in causally-interpretable
meta-analysis to account for between-study heterogeneity beyond
covariate differences between trials and the target population. Our
causal framework attempts to bridge the structure of traditional
random-effects meta-analysis with causal inference. To do so, we
introduce novel estimands, estimates, and inferential procedures that
all explicitly reference the role that a trial's setting can play in
driving systematic differences between study-specific parameters.

The impact of between-study heterogeneity can severely limit the
interpretability of treatment effect estimates transported from a
collection of trials to the target population. Under such heterogeneity,
estimates derived from data pooled across distinct trials may reflect a
combination of the idiosyncrasies present in each trial setting. In many
contexts, such heterogeneity persists despite standardizing the trial
covariate distributions to that of a target population. For instance,
variation in treatment version could cause differences in the potential
outcomes we would expect to observe across studies, even after
conditioning on all effect-modifiers, as well as covariates relevant to
treatment selection or trial participation.

To address these issues, we begin by defining potential outcomes that
explicitly reference setting-specific heterogeneity and, subsequently,
construct estimators and inferential procedures that take this
additional variability into account. By allowing potential outcomes to
depend on both assigned treatment and the setting of that assignment, we
can disaggregate the effects of a trial from the trial's population.

Our work is a first step toward clarifying the dual influence of trial
setting and participants in the context of causally-interpretable
meta-analysis. With additional information about protocol-level
differences between trials, our framework might be extended to reference
specific trial characteristics of interest. Future work might consider
transporting or up-weighting a specific source of heterogeneity
(e.g.~differing levels of adherence) to produce transported effects most
relevant to the clinical setting of interest. Relatedly, it is also
important to clarify the distinction between intention-to-treat and
per-protocol effects when attempting to account for setting-specific
heterogeneity. That is, investigators should distinguish between
studying what would happen if a member of the target population had
participated in a given trial and was compliant or was simply assigned
treatment. The estimands introduced in this work should be interpreted
as defining intention-to-treat transported effects, as our observed data
contains information on treatment assignment alone. However, these
methods may be extended to describe per-protocol effects when such data
are available.

In part, our work follows from a simple acknowledgment that the
particular context of an RCT has relevance for causal interpretability.
In other words, estimates transported from RCTs do not necessarily have
a generic interpretation uncoupled from the effects of trial setting. We
view our framework as one step toward developing a causal structure that
can accommodate between-study heterogeneity. In the future, we aim to
expand on this structure to incorporate trial-specific information
relevant to a variety of clinical questions in populations of interest.

\hypertarget{references}{%
\section*{References}\label{references}}
\addcontentsline{toc}{section}{References}

\hypertarget{refs}{}
\begin{CSLReferences}{1}{0}
\leavevmode\hypertarget{ref-jama_meta_analysis}{}%
Berlin, Jesse A., and Robert M. Golub. 2014. {``Meta-Analysis as
{Evidence}: {Building} a {Better} {Pyramid}.''} \emph{JAMA} 312 (6):
603. \url{https://doi.org/10.1001/jama.2014.8167}.

\leavevmode\hypertarget{ref-dahabreh_epi}{}%
Dahabreh, Issa J., Lucia C. Petito, Sarah E. Robertson, Miguel A.
Hernán, and Jon A. Steingrimsson. 2020. {``Toward {Causally}
{Interpretable} {Meta}-Analysis: {Transporting} {Inferences} from
{Multiple} {Randomized} {Trials} to a {New} {Target} {Population}.''}
\emph{Epidemiology} 31 (3): 334--44.
\url{https://doi.org/10.1097/EDE.0000000000001177}.

\leavevmode\hypertarget{ref-dahabreh_adherence}{}%
Dahabreh, Issa J., Sarah E. Robertson, and Miguel A. Hernán. 2022.
{``Generalizing and Transporting Inferences about the Effects of
Treatment Assignment Subject to Non-Adherence.''}
\url{https://arxiv.org/abs/2211.04876}.

\leavevmode\hypertarget{ref-dahabreh_biometrics}{}%
Dahabreh, Issa J., Sarah E. Robertson, Lucia C. Petito, Miguel A.
Hernán, and Jon A. Steingrimsson. 2022. {``Efficient and Robust Methods
for Causally Interpretable Meta-Analysis: Transporting Inferences from
Multiple Randomized Trials to a Target Population.''} \emph{Biometrics}.
https://doi.org/\url{https://doi.org/10.1111/biom.13716}.

\leavevmode\hypertarget{ref-dahabreh_scale_up_preprint}{}%
Dahabreh, Issa J., James M. Robins, Sebastien J.-P. A. Haneuse, and
Miguel A. Hernán. 2019. {``Generalizing Causal Inferences from
Randomized Trials: Counterfactual and Graphical Identification.''}
arXiv. \url{http://arxiv.org/abs/1906.10792}.

\leavevmode\hypertarget{ref-generalizability}{}%
Degtiar, Irina, and Sherri Rose. 2021. {``A Review of Generalizability
and Transportability.''}
\url{https://doi.org/10.48550/ARXIV.2102.11904}.

\leavevmode\hypertarget{ref-dersimonian}{}%
DerSimonian, Rebecca, and Nan Laird. 1986. {``Meta-Analysis in Clinical
Trials.''} \emph{Controlled Clinical Trials} 7 (3): 177--88.

\leavevmode\hypertarget{ref-cms_rct_variation}{}%
Dhruva, Sanket S. 2008. {``Variations {Between} {Clinical} {Trial}
{Participants} and {Medicare} {Beneficiaries} in {Evidence} {Used} for
{Medicare} {National} {Coverage} {Decisions}.''} \emph{Archives of
Internal Medicine} 168 (2): 136.
\url{https://doi.org/10.1001/archinternmed.2007.56}.

\leavevmode\hypertarget{ref-sort_taxonomy}{}%
Ebell, Mark H, Jay Siwek, Barry D Weiss, Steven H Woolf, Jeffrey Susman,
Bernard Ewigman, and Marjorie Bowman. 2004. {``Strength of
Recommendation Taxonomy (SORT): A Patient-Centered Approach to Grading
Evidence in the Medical Literature.''} \emph{The Journal of the American
Board of Family Practice} 17 (1): 59--67.

\leavevmode\hypertarget{ref-rwe_nam}{}%
Galson, Steven, and Gregory Simon. 2016. {``Real-{World} {Evidence} to
{Guide} the {Approval} and {Use} of {New} {Treatments}.''} \emph{NAM
Perspectives} 6 (10). \url{https://doi.org/10.31478/201610b}.

\leavevmode\hypertarget{ref-re_meta}{}%
Higgins, Julian P. T., Simon G. Thompson, and David J. Spiegelhalter.
2009. {``A Re-Evaluation of Random-Effects Meta-Analysis.''}
\emph{Journal of the Royal Statistical Society. Series A (Statistics in
Society)} 172 (1): 137--59. \url{http://www.jstor.org/stable/30136745}.

\leavevmode\hypertarget{ref-panns_general}{}%
Kay, Stanley R., Abraham Fiszbein, and Lewis A. Opler. 1987. {``{The
Positive and Negative Syndrome Scale (PANSS) for Schizophrenia}.''}
\emph{Schizophrenia Bulletin} 13 (2): 261--76.
\url{https://doi.org/10.1093/schbul/13.2.261}.

\leavevmode\hypertarget{ref-clinical_practice_bad}{}%
Lunny, Carole, Cynthia Ramasubbu, Lorri Puil, Tracy Liu, Savannah
Gerrish, Douglas M. Salzwedel, Barbara Mintzes, and James M. Wright.
2021. {``Over Half of Clinical Practice Guidelines Use Non-Systematic
Methods to Inform Recommendations: {A} Methods Study.''} Edited by Tim
Mathes. \emph{PLOS ONE} 16 (4): e0250356.
\url{https://doi.org/10.1371/journal.pone.0250356}.

\leavevmode\hypertarget{ref-if_wild_bootstrap}{}%
Matsouaka, Roland A., Yi Liu, and Yunji Zhou. 2022. {``Variance
Estimation for the Average Treatment Effects on the Treated and on the
Controls.''} arXiv. \url{https://doi.org/10.48550/ARXIV.2209.10742}.

\leavevmode\hypertarget{ref-tau_paper}{}%
Rao, Poduri S. R. S., Jack Kaplan, and William G. Cochran. 1981.
{``Estimators for the One-Way Random Effects Model with Unequal Error
Variances.''} \emph{Journal of the American Statistical Association} 76
(373): 89--97. \url{http://www.jstor.org/stable/2287050}.

\leavevmode\hypertarget{ref-monoclonal_antibody}{}%
Stadler, Eva, Khai Li Chai, Timothy E Schlub, Deborah Cromer, Mark N
Polizzotto, Stephen J Kent, Nicole Skoetz, et al. 2022. {``Determinants
of Passive Antibody Effectiveness in SARS-CoV-2 Infection.''}
\emph{medRxiv}. \url{https://doi.org/10.1101/2022.03.21.22272672}.

\leavevmode\hypertarget{ref-tsiatis_semiparametric}{}%
Tsiatis, Anastasios A. 2006. {``Semiparametric Theory and Missing
Data.''}

\leavevmode\hypertarget{ref-fda_guidance}{}%
U.S. Food and Drug Administration. 2021. {``FDA Guidance on Conduct of
Clinical Trials of Medical Products During the COVID-19 Public Health
Emergency.''} Guidance for Industry, Investigators, and Institutional
Review Boards FDA-2020-D-1106-0002.

\leavevmode\hypertarget{ref-ipd_meta_uptake}{}%
Vale, C. L., L. H. M. Rydzewska, M. M. Rovers, J. R. Emberson, F.
Gueyffier, L. A. Stewart, and on behalf of the Cochrane IPD
Meta-analysis Methods Group. 2015. {``Uptake of Systematic Reviews and
Meta-Analyses Based on Individual Participant Data in Clinical Practice
Guidelines: Descriptive Study.''} \emph{BMJ} 350 (mar06 6): h1088--88.
\url{https://doi.org/10.1136/bmj.h1088}.

\leavevmode\hypertarget{ref-mult_versions}{}%
VanderWeele, Tyler J., and Miguel A. Hernán. 2013. {``Causal Inference
Under Multiple Versions of Treatment.''} \emph{Journal of Causal
Inference} 1 (1): 1--20.
\url{https://doi.org/doi:10.1515/jci-2012-0002}.

\end{CSLReferences}

\hypertarget{acknowledgements}{%
\section*{Acknowledgements}\label{acknowledgements}}
\addcontentsline{toc}{section}{Acknowledgements}

The authors acknowledge the Minnesota Supercomputing Institute (MSI) at
the University of Minnesota for providing resources that contributed to
the research results reported within this paper. URL:
\url{http://www.msi.umn.edu}

\newpage

\hypertarget{appendix}{%
\section{Appendix}\label{appendix}}

\hypertarget{full-proof-of-identification-formula-result-for-mu_a-0ka_s}{%
\subsection{\texorpdfstring{Full proof of identification formula result
for
\(\mu_{a, 0}(k^a_s)\)}{Full proof of identification formula result for \textbackslash mu\_\{a, 0\}(k\^{}a\_s)}}\label{full-proof-of-identification-formula-result-for-mu_a-0ka_s}}

\begin{align*}
    E[Y(a, k^a_s)|R=0] &= E[E[Y(a, k^a_s)|X, R=0]|R=0] & \text{Law of total expectation}\\
    &= E[E[Y(a, k^a_s)|X, S=s]|R=0] & \text{By Assumption 1}\\ &= E[E[Y(a, k^a_s)|X, S=s, A=a]|R=0] & \text{By Assumption 2}\\
    &= E[E[Y(a, k^a_s)|X, S=s, A=a, K^a(a)=k^a_s]|R=0] & \text{By Assumption 5}\\
    &= E[E[Y|X, S=s, A=a]|R=0] & \text{By Assumption 3}
\end{align*}

\hypertarget{full-derivation-of-the-approximation-of-eya-ka_0ar0}{%
\subsection{\texorpdfstring{Full derivation of the approximation of
\(E[Y(a, K^a_0(a))|R=0]\)}{Full derivation of the approximation of E{[}Y(a, K\^{}a\_0(a))\textbar R=0{]}}}\label{full-derivation-of-the-approximation-of-eya-ka_0ar0}}

\begin{align*}
    E[Y(a, K^a_0(a))|R=0] &= \sum_{k^a\in \mathcal{K}^a}E\left\{E[Y(a, K^a_0(a))|R=0, K^a_0(a)=k^a]|R=0\right\}P(K^a_0(a)=k^a)\\
    &\approx \sum_{k^a\in\mathcal{K}^a}E\left\{E[Y(a, K^a_0(a))|R=0, K^a_0(a)=k^a]|R=0\right\}\hat{P}(K^a_0(a)=k^a)\\
    &= \sum_{k^a\in\mathcal{K}^a}E\left\{E[Y(a, K^a_0(a))|R=0, K^a_0(a)=k^a]|R=0\right\}\left\{\frac{1}{m}\sum_{s=1}^m \mathbbm{1}(k^a=k^a_s)\right\}\\
    &= \sum_{k^a\in\mathcal{K}^a}\frac{1}{m}\sum_{s=1}^m \mathbbm{1}(k^a=k^a_s)E\left\{E[Y(a, K^a_s(a))|R=0, K^a_s(a)=k^a]|R=0\right\}\\
    &= \frac{1}{m}\sum_{s=1}^m\sum_{k^a\in\mathcal{K}^a} \mathbbm{1}(k^a=k^a_s)E\left\{E[Y(a, K^a_s(a))|R=0, K^a_s(a)=k^a]|R=0\right\}\\
    &= \frac{1}{m}\sum_{s=1}^mE\left\{E[Y(a, K^a_s(a))|R=0, K^a_s(a)=k^a_s]|R=0\right\}\\
    &= \frac{1}{m}\sum_{s=1}^m E[Y(a, k^a_s)|R=0]\\
    &= \frac{1}{m}\sum_{s=1}^m E[E[Y|X, S=s, A=a]|R=0] \hspace{3.4cm} \text{By \eqref{eq: id_one_trial}}
\end{align*}

\hypertarget{full-derivation-of-our-estimate-of-between-study-variance-in-the-presence-of-correlated-estimates}{%
\subsection{Full derivation of our estimate of between-study variance in
the presence of correlated
estimates}\label{full-derivation-of-our-estimate-of-between-study-variance-in-the-presence-of-correlated-estimates}}

Applying the notation defined in the main manuscript, write the overall
sum of squares of our estimators as \begin{equation*}
    Q=\sum_{s=1}^m \left(\hat{\mu}_s-\hat{\mu}\right)^2 = \sum_{s=1}^m \left(\hat{\mu}_s-\sum_{l=1}^m w_l\hat{\mu}_l\right)^2
\end{equation*} and \begin{equation*}
    Q_s = \left(\hat{\mu}_s-\hat{\mu}\right)^2 = \left(\hat{\mu}_s-\sum_{l=1}^m w_l\hat{\mu}_l\right)^2.
\end{equation*} Thus, \(Q=\sum_{s=1}^m Q_s\). As in Rao, Kaplan, and
Cochran (\protect\hyperlink{ref-tau_paper}{1981}), we derive an
expression for the expectation of \(Q\) and an estimator for
\(\gamma^2\). To that end, let \begin{equation*}
    \mathbf{a}_s = \begin{pmatrix} -w_1\\
    \vdots\\
    1-w_s\\
    \vdots\\
    -w_m\end{pmatrix}
\end{equation*} and \begin{equation*}
    \Sigma =
  \begin{pmatrix}
    \sigma^2_1 + \gamma^2 & \sigma_{12} & \cdots & \sigma_{1m} \\
    \sigma_{12} & \sigma^2_2 + \gamma^2 & \cdots & \sigma_{2m}\\
    \vdots & \vdots & \vdots & \vdots\\
    \sigma_{1m} & \sigma_{2m} & \cdots & \sigma^2_m + \gamma^2
  \end{pmatrix}
\end{equation*} be the variance-covariance matrix for
\(\hat{\pmb{\mu}}\) with \(\sigma^2_i+\gamma^2\) as the \(i\)th diagonal
entry and \(\sigma_{ij}\) as the \((i, j)\) off diagonal entry
(i.e.~\(\sigma_{ij}\) is \(\text{Cov}(\hat{\mu}_i, \hat{\mu}_j)\)).
Then, \begin{align*}
    \mathbf{a}_s\mathbf{a}_s^T\Sigma &= \begin{pmatrix} -w_1\\
    \vdots\\
    1-w_s\\
    \vdots\\
    -w_m\end{pmatrix}\begin{pmatrix} -w_1 & \cdots & 1-w_s & \cdots -w_m\end{pmatrix}\begin{pmatrix}\sigma^2_1 + \gamma^2 & \sigma_{12} & \cdots & \sigma_{1m} \\
    \sigma_{12} & \sigma^2_2 + \gamma^2 & \cdots & \sigma_{2m}\\
    \vdots & \vdots & \vdots & \vdots\\
    \sigma_{1m} & \sigma_{2m} & \cdots & \sigma^2_m + \gamma^2
  \end{pmatrix}\\
  &= \begin{pmatrix} w_1^2 & \cdots & -w_1(1-w_s) & \cdots &  w_1w_m\\
  \vdots & \vdots & \vdots & \vdots & \vdots\\
  -w_1(1-w_s) & \cdots & (1-w_s)^2 & \cdots & -w_m(1-w_s)\\
  \vdots & \vdots & \vdots & \vdots & \vdots\\
  w_1w_m & \cdots & -w_m(1-w_s) & \cdots & w_m^2\end{pmatrix}\begin{pmatrix}\sigma^2_1 + \gamma^2 & \sigma_{12} & \cdots & \sigma_{1m} \\
    \sigma_{12} & \sigma^2_2 + \gamma^2 & \cdots & \sigma_{2m}\\
    \vdots & \vdots & \vdots & \vdots\\
    \sigma_{1m} & \sigma_{2m} & \cdots & \sigma^2_m + \gamma^2
  \end{pmatrix}.
\end{align*} That is, after computing \(\mathbf{a}_s\mathbf{a}_s^T\), we
obtain a matrix where the \(i\)th diagonal entry is \(w_i^2\) except for
the \(s\)th such entry which is \((1-w_s)^2\). In each row \(i\),
\(i\neq s\), entries \((i, j)\), \(j\neq s\) and \(j\neq i\) are given
by \(w_iw_j\) while entry \((i, s)\) is \(-w_i(1-w_s)\). In the \(s\)th
row, each entry \((s, j)\), \(j\neq s\) is \(-w_j(1-w_s)\). To compute
\(E[Q_s]\), we need to compute
\(\text{tr}(\mathbf{a}_s\mathbf{a}_s^T\Sigma)\). Looking at the matrices
in the second equality above, taking the dot product of the first row in
the matrix of weights with the first column of \(\Sigma\) we can see
that the \(i\)th such diagonal entry (\(i\neq s\)) \begin{equation*}
    w_i^2(\sigma^2_i+\gamma^2) -w_i(1-w_s)(\sigma_{is})+\sum_{j\neq i, j\neq s}w_iw_j\sigma_{ij}.
\end{equation*} The \(s\)th diagonal entry is then \begin{equation*}
    (1-w_s)^2(\sigma^2_s+\gamma^2) + \sum_{i\neq s}(-w_i)(1-w_s)\sigma_{is}.
\end{equation*} Summing all the diagonal entries, \begin{align*}
    \text{tr}(\mathbf{a}_s\mathbf{a}_s^T\Sigma) &= (1-w_s)^2(\sigma^2_s+\gamma^2) + \sum_{i\neq s}(-w_i)(1-w_s)\sigma_{is} + \sum_{i\neq s}\left[w_i^2(\sigma^2_i+\gamma^2) -w_i(1-w_s)(\sigma_{is})+\sum_{j\neq i, j\neq s}w_iw_j\sigma_{ij}\right]\\
    &= (1-2w_s)(\sigma^2_s+\gamma^2)+\sum_{i=1}^m w_i^2(\sigma^2_i+\gamma^2)-2\sum_{i\neq s}w_i(1-w_s)\sigma_{is}+\sum_{i\neq s}\left[\sum_{j\neq i, j\neq s}w_iw_j\sigma_{ij}\right].
\end{align*} Now the question is: how can we compute \(\gamma^2\) using
the moment equations? To begin, let \begin{equation*}
    C_s = -2\sum_{i\neq s}w_i(1-w_s)\sigma_{is}+\sum_{i\neq s}\left[\sum_{j\neq i, j\neq s}w_iw_j\sigma_{ij}\right]
\end{equation*} denote all the components of
\(\text{tr}(\mathbf{a}_s\mathbf{a}_s^T\Sigma)\) that are constant as a
function of \(\gamma^2\). Then, recognizing that
\(\pmb{\mu}^T\mathbf{a}_i\mathbf{a}_i^T\pmb{\mu}=0\), we have
\begin{equation*}
    E[Q_s] = (1-2w_s)(\sigma^2_s+\gamma^2)+\sum_{i=1}^m w_i^2(\sigma^2_i+\gamma^2)+C_s
\end{equation*} and, moreover, \begin{align*}
    E[Q]&=\sum_{s=1}^m E[Q_s]\\
    &= \sum_{s=1}^m \left[(1-2w_s)(\sigma^2_s+\gamma^2)+\sum_{i=1}^m w_i^2(\sigma^2_i+\gamma^2)+C_s\right]\\
    &= m\sum_{s=1}^m w_s^2(\sigma^2_s+\gamma^2)+\sum_{s=1}^m \left[(1-2w_s)(\sigma^2_s+\gamma^2)+C_s\right].
\end{align*} We apply this expression for \(E[Q]\) to derive the
estimate for between-study heterogeneity given in the main part of the
paper.

\newpage

\hypertarget{identification-for-ipw-estimator}{%
\subsection{Identification for IPW
estimator}\label{identification-for-ipw-estimator}}

This identification result proceeds similarly to the analogous IPW
identification result in Dahabreh et al.
(\protect\hyperlink{ref-dahabreh_biometrics}{2022}). The important
distinction is that the weights are distinct for each trial, rather than
applying to all members of the trial population. The result below
follows from application of iterated expectation and our positivity
assumption (Assumption 4).

\begin{align*}
    \psi_{s, 0}(a) &= E[E[Y|X, S=s, A=a]|R=0]\\
    &= E\left.\left[E\left[\left.\frac{I(S=s, A=a)Y}{P(S=s|X)P(A=a|X, S=s)}\right\vert X\right]\right\vert R=0\right]\\
    &= \frac{1}{P(R=0)}E\left[I(R=0)E\left.\left[\frac{I(S=s, A=a)Y}{P(S=s|X)P(A=a| X, S=s)}\right\vert X\right]\right]\\
    &= \frac{1}{P(R=0)}E\left[E\left.\left[\frac{I(S=s, A=a)YP(R=0|X)}{P(S=s|X)P(A=a|X, S=s)}\right \vert X\right]\right]\\
    &= \frac{1}{P(R=0)}E\left[E\frac{I(S=s, A=a)YP(R=0|X)}{P(S=s|X)P(A=a|X, S=s)}\right]
\end{align*}

\hypertarget{full-simulation-results-quantile-based-prediction-intervals}{%
\subsection{Full Simulation Results: Quantile-Based Prediction
Intervals}\label{full-simulation-results-quantile-based-prediction-intervals}}

\includegraphics{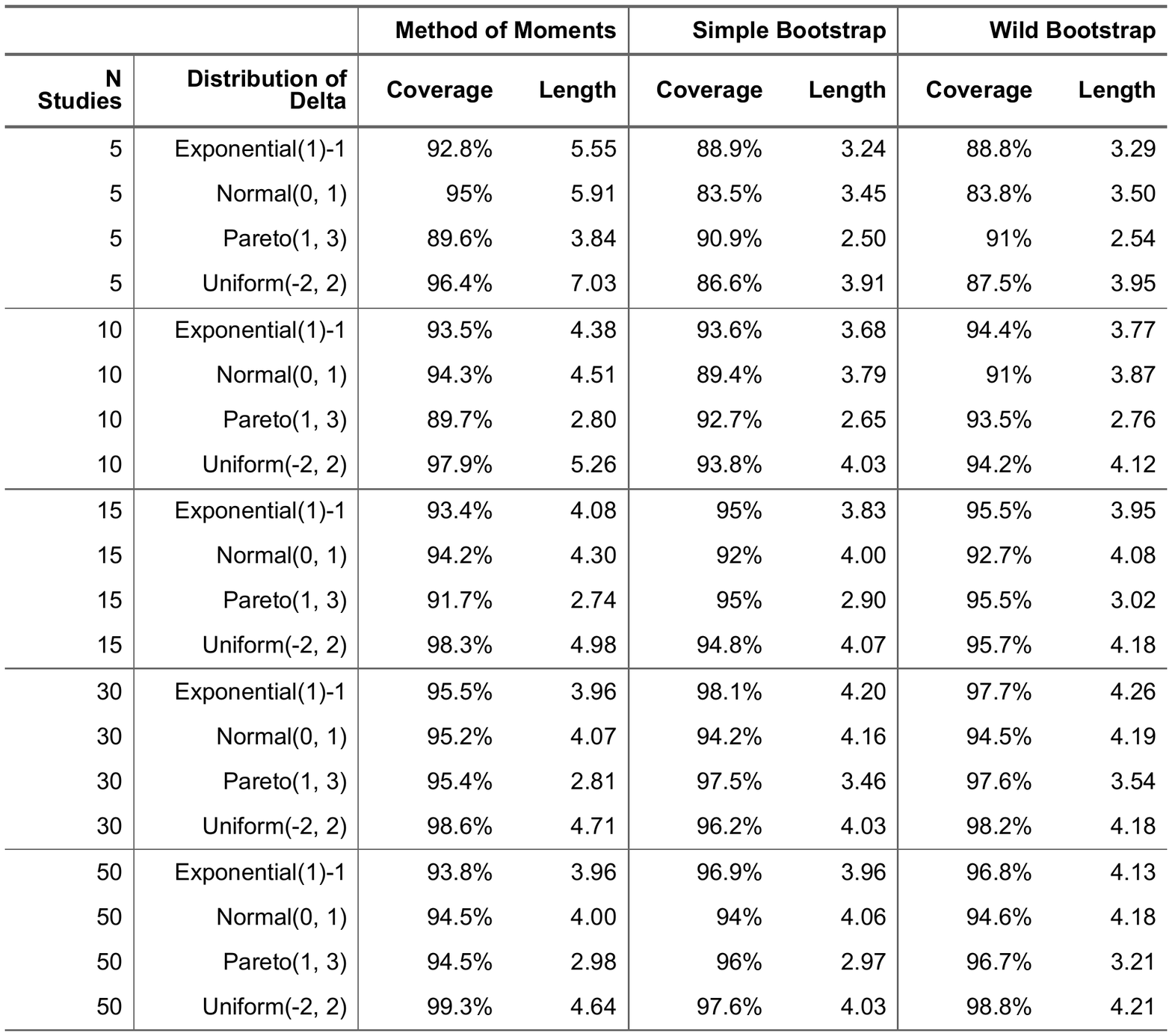}

\newpage

\hypertarget{full-simulation-results-normal-based-prediction-intervals}{%
\subsection{Full Simulation Results: Normal-Based Prediction
Intervals}\label{full-simulation-results-normal-based-prediction-intervals}}

\includegraphics{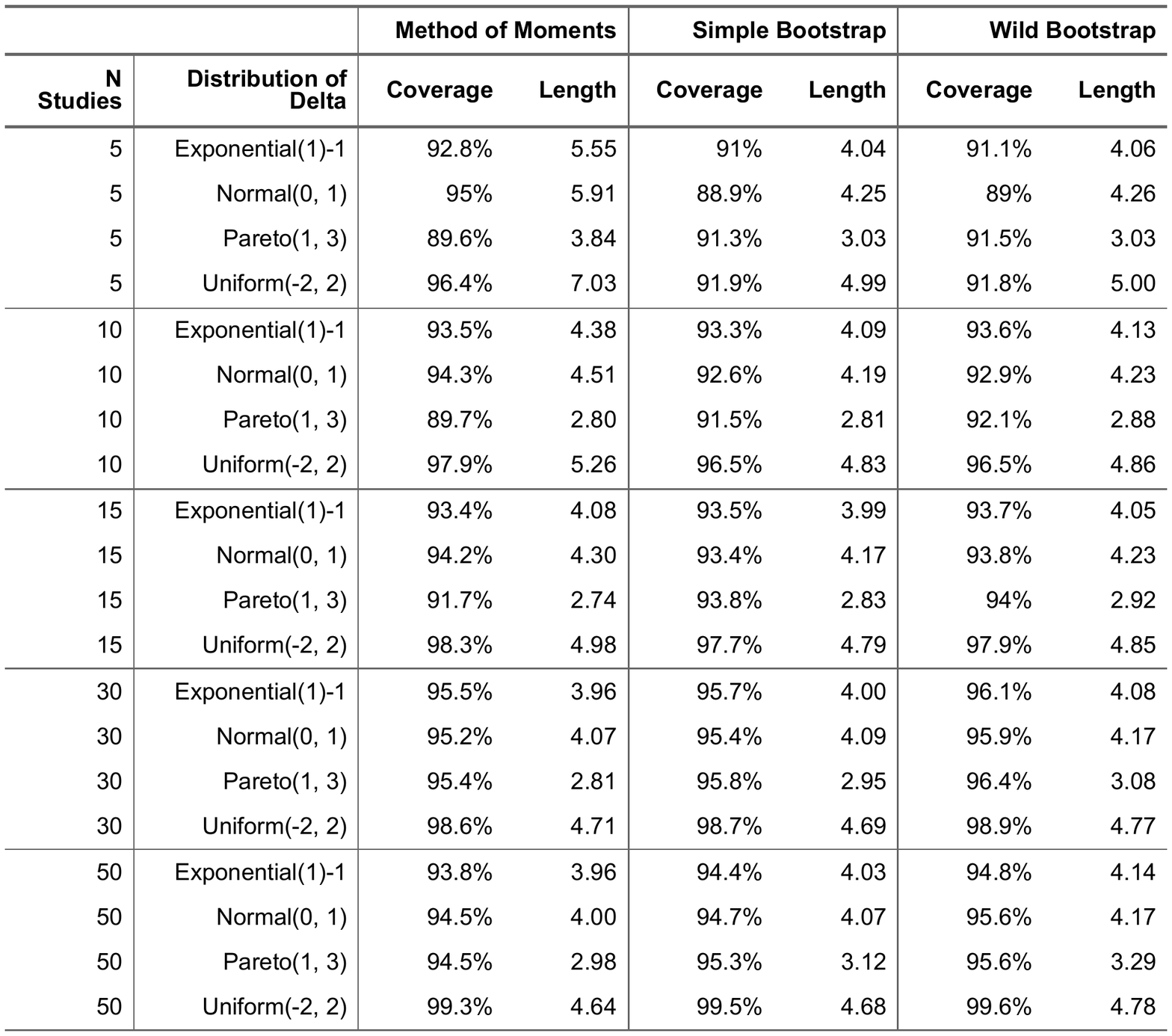}

\end{document}